\documentclass[%
 reprint,
superscriptaddress,
 preprintnumbers,
 amsmath,amssymb,
 aps,
 prl,
]{revtex4-1}

\usepackage{color}

\usepackage[colorlinks=true,linkcolor=blue,citecolor=blue, urlcolor=blue]{hyperref}   %

\usepackage{graphicx}%
\usepackage{dcolumn}%
\usepackage{bm}%
\usepackage{xcolor}
\def\Fig#1{Fig.~\ref{#1}}

\def\trento{T\raisebox{-0.5ex}{R}ENTo model}

\begin{document}
\title{A transverse momentum differential global analysis of Heavy Ion Collisions}
\author{Govert Nijs}
\email{govert@mit.edu}
\affiliation{Center for Theoretical Physics, Massachusetts Institute of Technology, Cambridge, MA 02139, USA}
\affiliation{Institute for Theoretical Physics and Center for Extreme Matter and Emergent Phenomena, Utrecht University, 3584 CC Utrecht, The Netherlands}
\author{Wilke van der Schee}
\email{wilke.van.der.schee@cern.ch}
\affiliation{Theoretical Physics Department, CERN, CH-1211 Gen\`eve 23, Switzerland}
\author{Umut G\"ursoy}
\email{u.gursoy@uu.nl}
\affiliation{Institute for Theoretical Physics and Center for Extreme Matter and Emergent Phenomena, Utrecht University, 3584 CC Utrecht, The Netherlands}
\author{Raimond Snellings}
\email{raimond.snellings@nikhef.nl}
\affiliation{Institute for Gravitational and Subatomic Physics (GRASP), Utrecht University, 3584 CC Utrecht, The Netherlands}
\affiliation{Nikhef, 1098 XG Amsterdam, The Netherlands}
\begin{abstract}
The understanding of heavy ion collisions and its quark-gluon plasma (QGP) formation requires a complicated interplay of rich physics in a wealth of experimental data. In this work we compare for identified particles the transverse momentum dependence of both the yields and the anisotropic flow coefficients for both PbPb and $p$Pb collisions. We do this in a global model fit including a free streaming prehydrodynamic phase with variable velocity $v_\text{fs}$, thereby widening the scope of initial conditions. During the hydrodynamic phase we  vary three second order transport coefficients. The free streaming velocity has a preference slightly below the speed of light. In this extended model the QGP bulk viscosity is small and even consistent with zero.
\end{abstract}
\preprint{CERN-TH-2020-174//MIT-CTP/5250}

\maketitle

\noindent
{\bf Introduction -}
The quark-gluon plasma (QGP) is a state of deconfined matter of quarks and gluons whose existence at high energy density is predicted by Quantum Chromodynamics. Heavy ion collisions (HIC) at RHIC and LHC have led to a wealth of data from which the formation of this quark-gluon plasma can be inferred \cite{Heinz:2013th,Busza:2018rrf}. 
This existence can be deduced by having a model of initial conditions directly after the collision, a hydrodynamic phase with certain transport properties and lastly a hadronic phase of which the results can be compared to experimental results.
Even though robust conclusions on the existence of a QGP can be reached from qualitative features of the data, such as the anisotropy of the low transverse momentum particles or the quenching of high momentum partons, for a quantitative understanding of fundamental properties such as e.g.~the shear and bulk viscosity it is paramount to have a careful understanding of all parameters involved in all initial, hydrodyanamic and hadronic phases.

Early studies performing such a comprehensive analysis where all parameters in all stages can be varied simultaneously include \cite{Novak:2013bqa, Pratt:2015zsa, Sangaline:2015isa, Bernhard:2016tnd, Bernhard:2019bmu, Devetak:2019lsk, Auvinen:2020mpc}. This is done in similarity to modelling in cosmology \cite{Habib:2007ca}, where cosmological parameters have to be inferred from the Cosmic Microwave Background as well as Large Scale Structure analysis. Full simulations themselves are computationally expensive, and hence an emulator trained on few carefully selected design points is used to evaluate the likelihood of parameters using a Markov Chain Monte Carlo (mcmc). This, together with typically flat prior probability distributions, leads to final (Bayesian) posterior distributions for the chosen parameters.

For a precision study of QGP properties the scope of the full model is important, as artificially restricting e.g.~the range of initial conditions may pose unphysical restrictions on hydrodynamic transport.
In this Letter we will present the widest set of initial conditions studied to-date combined with hydrodynamics with varying second order transport coefficients, containing a total of 21 varying parameters (boldface in this Letter)\@. Most importantly, we perform a global analysis including experimental data with  transverse momentum dependence as well as particle identification for spectra and anisotropic flow coefficients for PbPb collisions at  2.76 and 5.02 TeV, together with identified spectra for $p$Pb collisions at 5.02 TeV\@.

\noindent
{\bf Model -}
For the initial conditions we use the \trento{} parametrization \cite{Moreland:2014oya,Moreland:2018gsh}. In this model nucleons of Gaussian width $\bf{w}$ are positioned by a fluctuating Glauber model separated by a distance of at least $\mathbf{d}_\text{min}$, and all nucleons consist of $\mathbf{n}_c$ %
randomly placed constituents having a Gaussian width of $v_\text{min} + \boldsymbol{\chi}_\text{struct} (\textbf{w}-v_\text{min})$, with $v_\text{min} = 0.2\,\text{fm}$\@. Nucleons are wounded depending on their overlap such that the cross section matches the proton-proton result. Constituents of wounded nucleons contribute to the left and right thickness functions $\mathcal{T}_A$ and $\mathcal{T}_B$ with norm $\mathbf{N}\gamma/\mathbf{n}_c$, where $\gamma$ fluctuates according to a gamma distribution of width $\boldsymbol{\sigma}_{\rm fluct}\sqrt{\mathbf{n}_c}$. These function are finally combined to a final parton density by $\mathcal{T}=(\tfrac{1}{2}\mathcal{T}_A^{\bf{p}}+\tfrac{1}{2} \mathcal{T}_B^{\bf{p}})^{1/\bf{p}}$ with $\bf{p}$ a free parameter.

The prehydrodynamic evolution consists of a free-streaming phase lasting for a time $\boldsymbol{\tau}_{\rm fs}$, with the new feature of introducing an effective velocity $\bf{v_{\rm fs}}$ (see also the discussion).

\begin{figure*}[ht]
\includegraphics[width=0.99\textwidth]{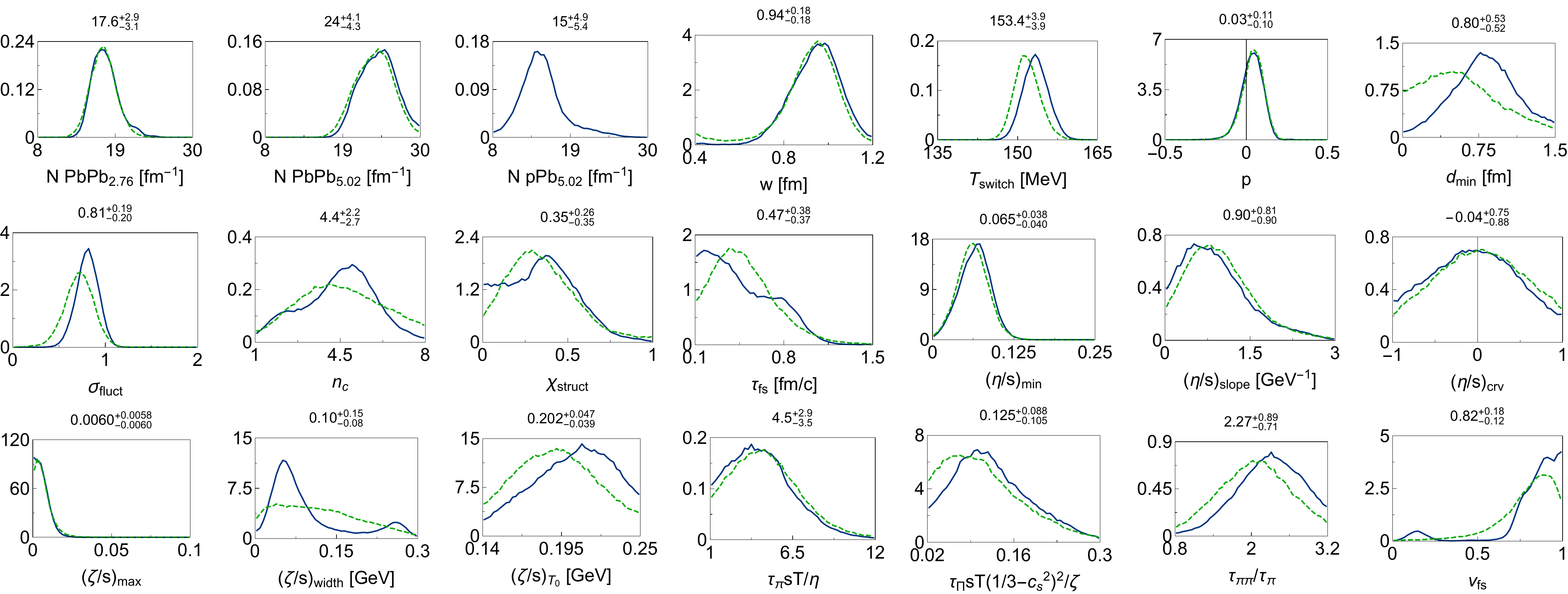}
\caption{\label{fig:posterior1d}Posterior distributions for all model parameters fitted to PbPb and $p$Pb (solid) or PbPb only (dashed, not applicable to $p$Pb norm) data. Values indicate the expectation values with the 90\% highest posterior density credible interval.}
\end{figure*}

\begin{figure*}[ht]
\includegraphics[width=0.32\textwidth]{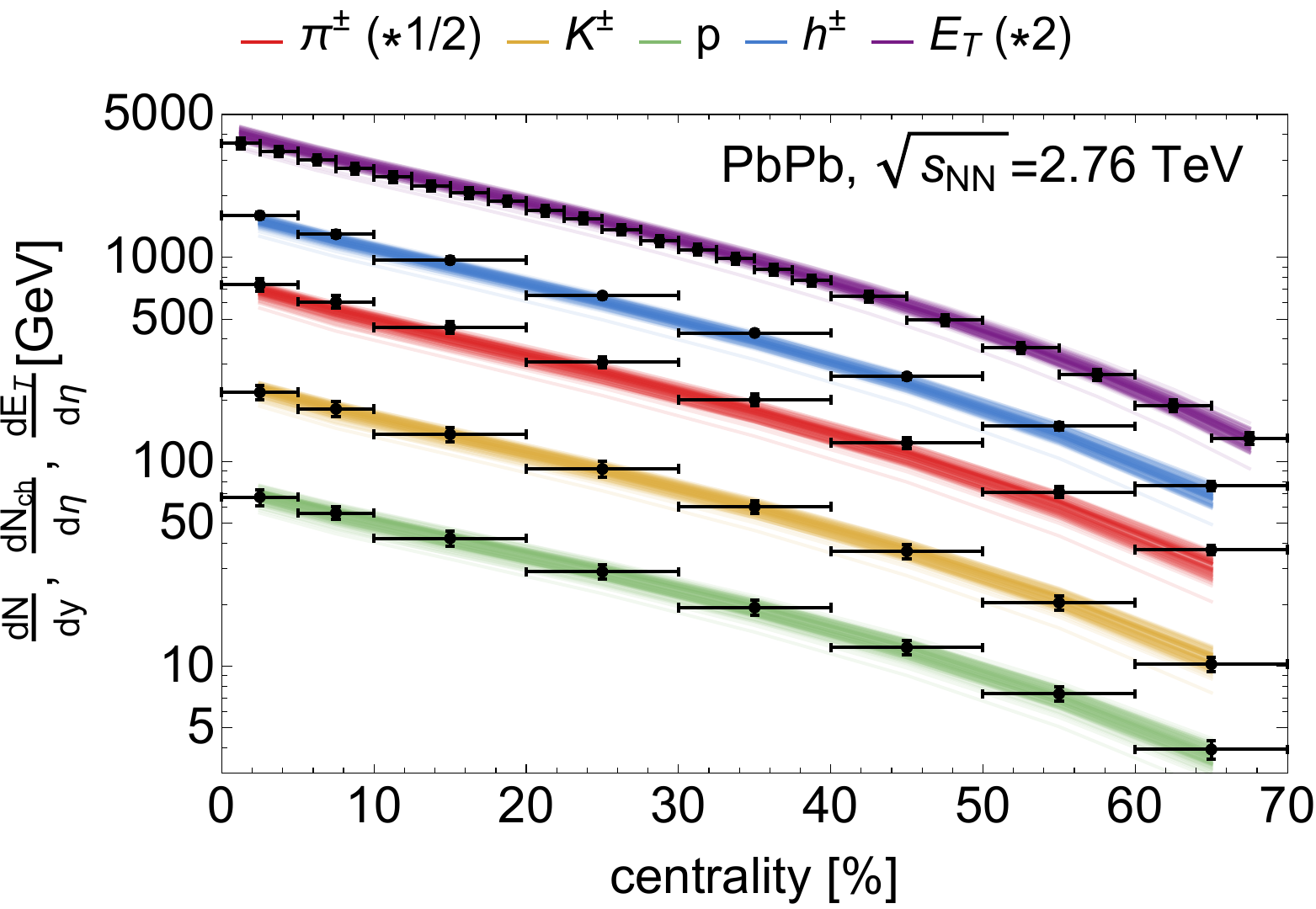}
\includegraphics[width=0.32\textwidth]{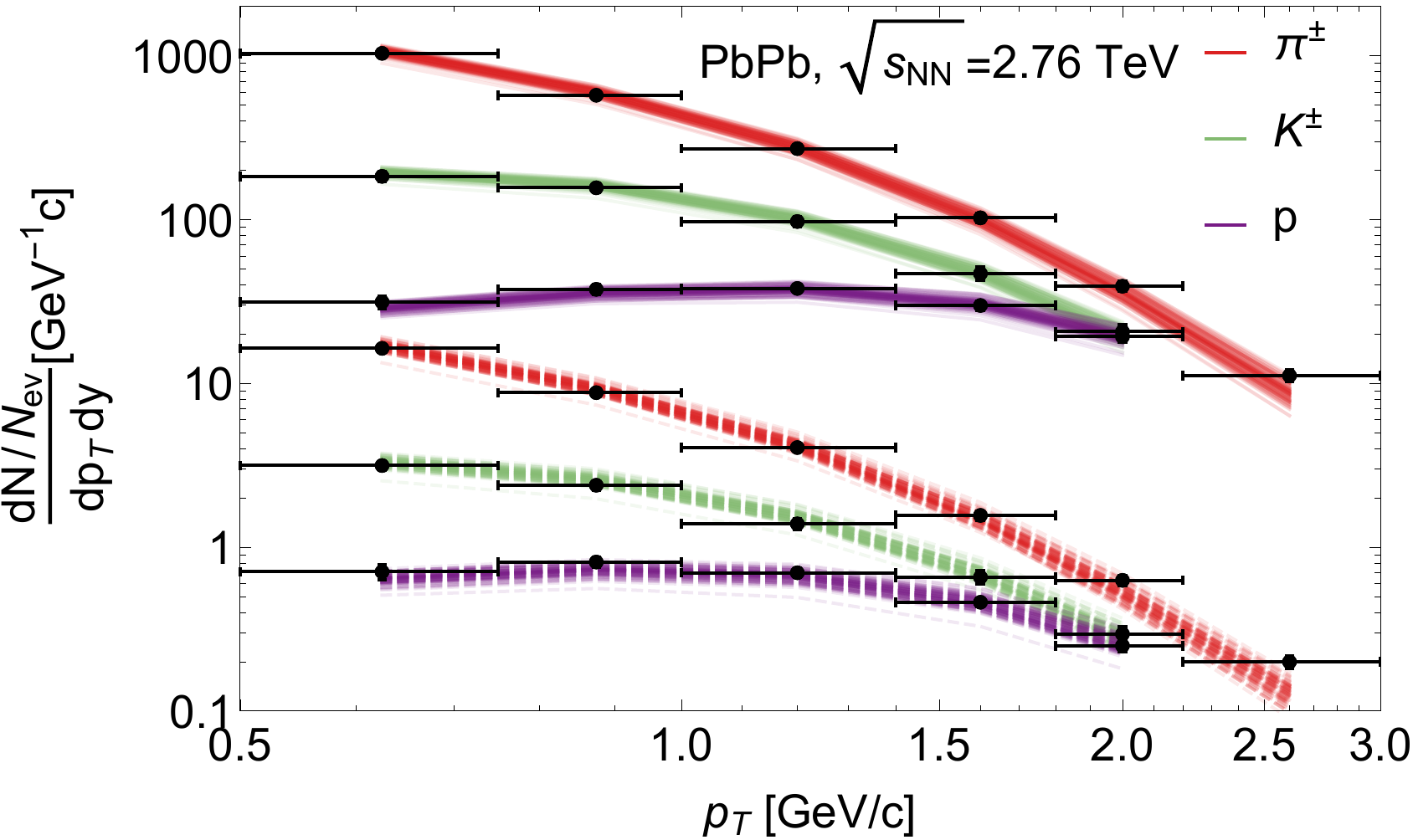}
\includegraphics[width=0.32\textwidth]{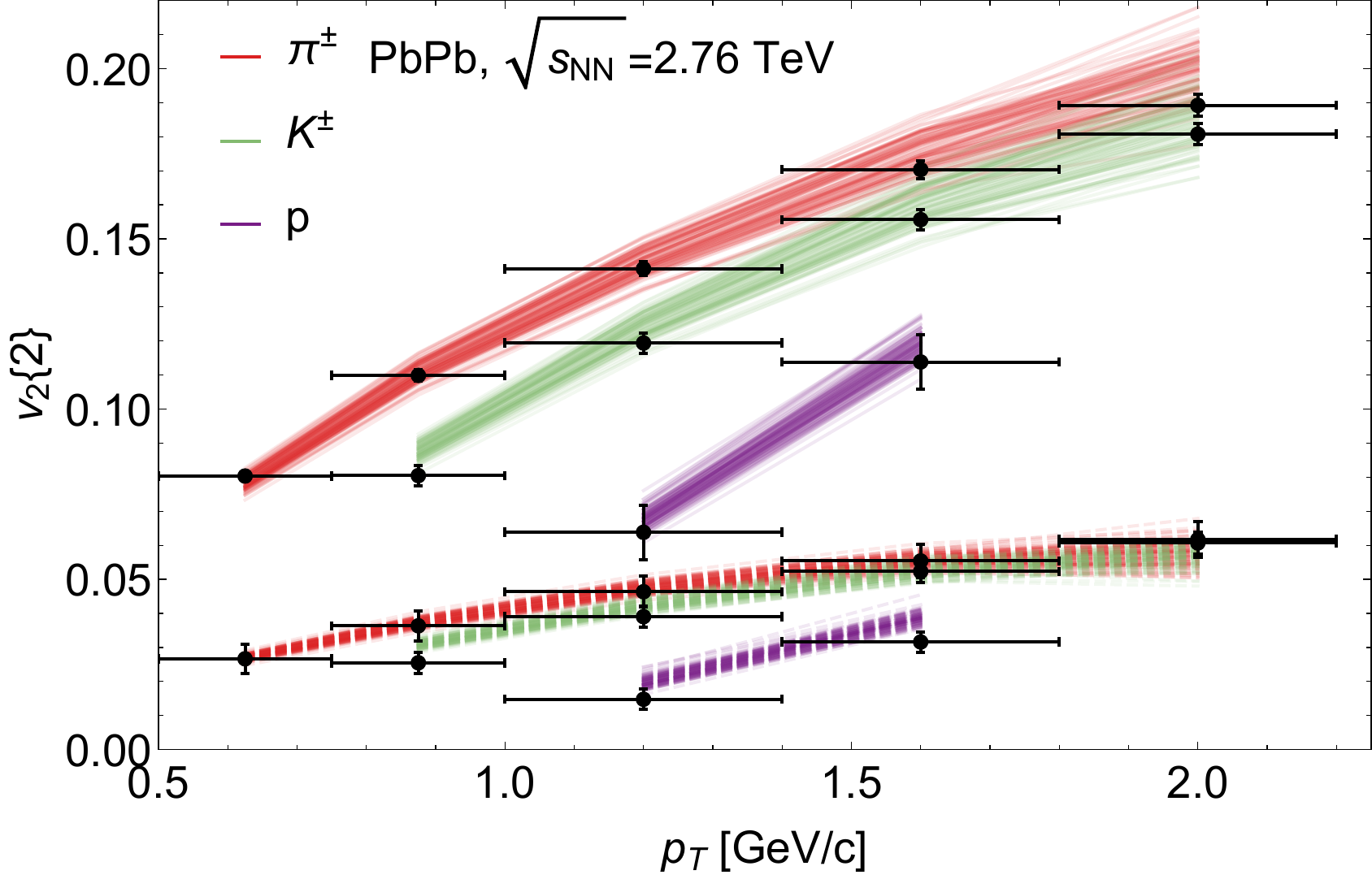}
\includegraphics[width=0.32\textwidth]{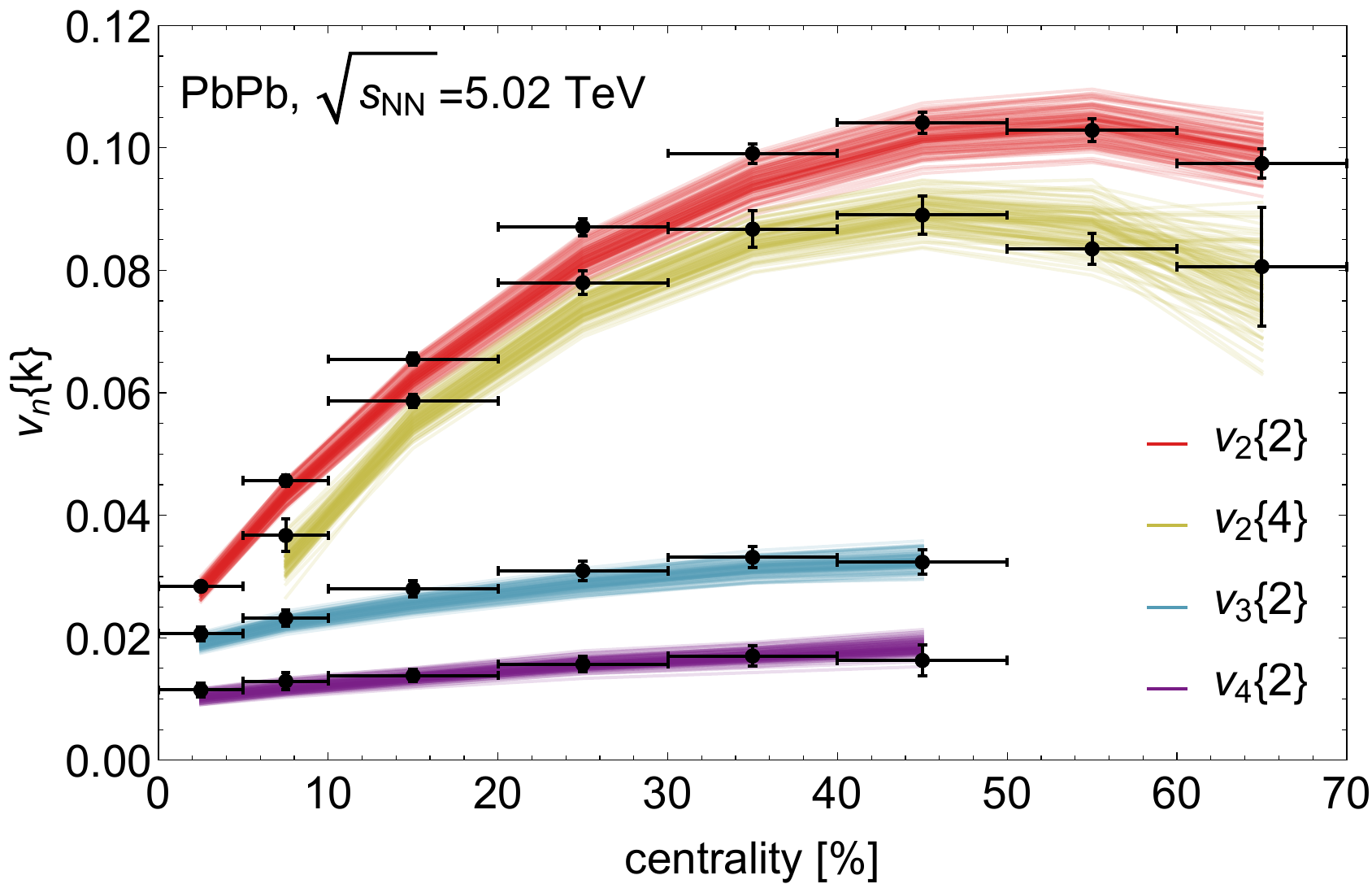}
\includegraphics[width=0.32\textwidth]{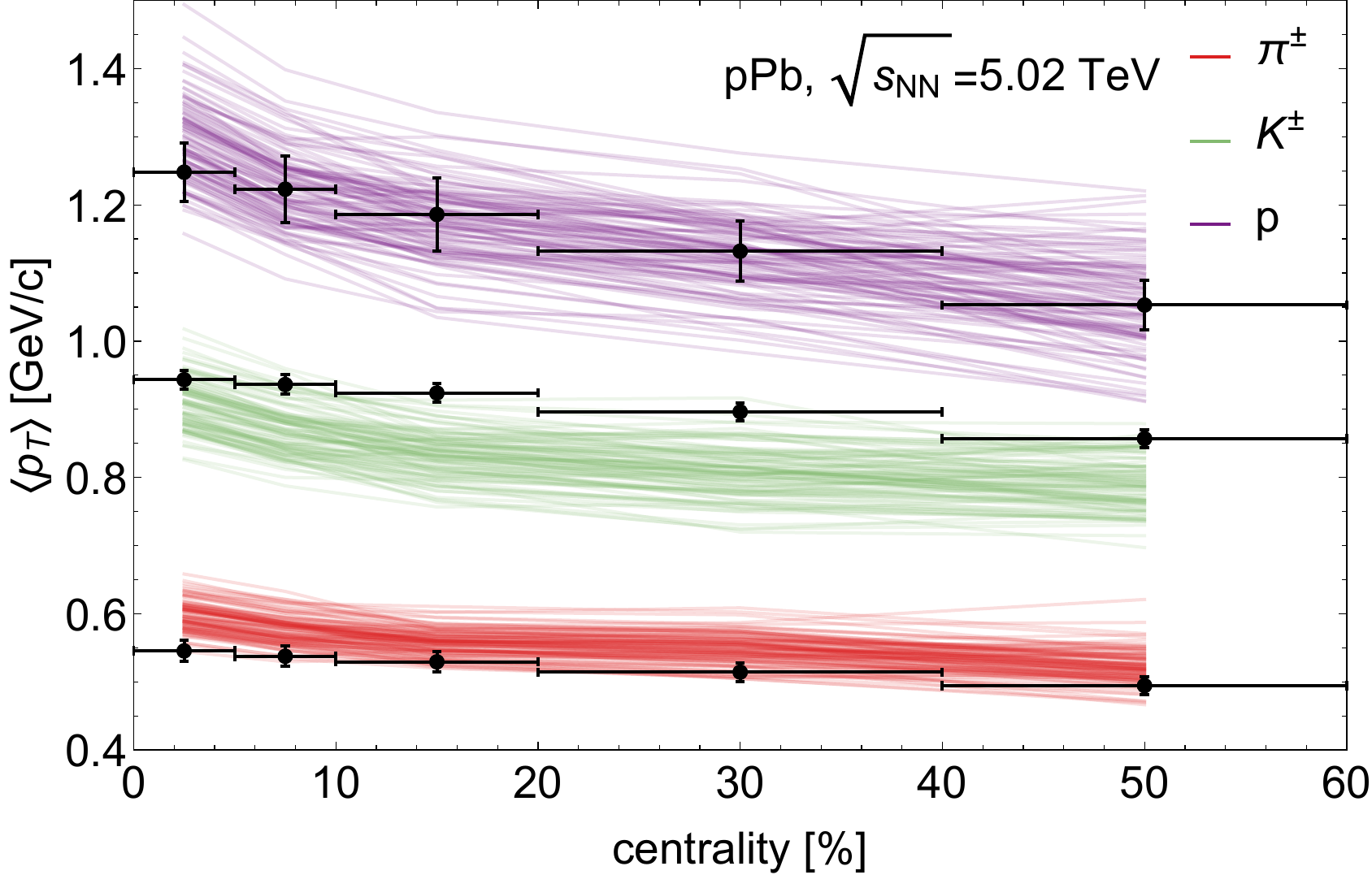}
\includegraphics[width=0.32\textwidth]{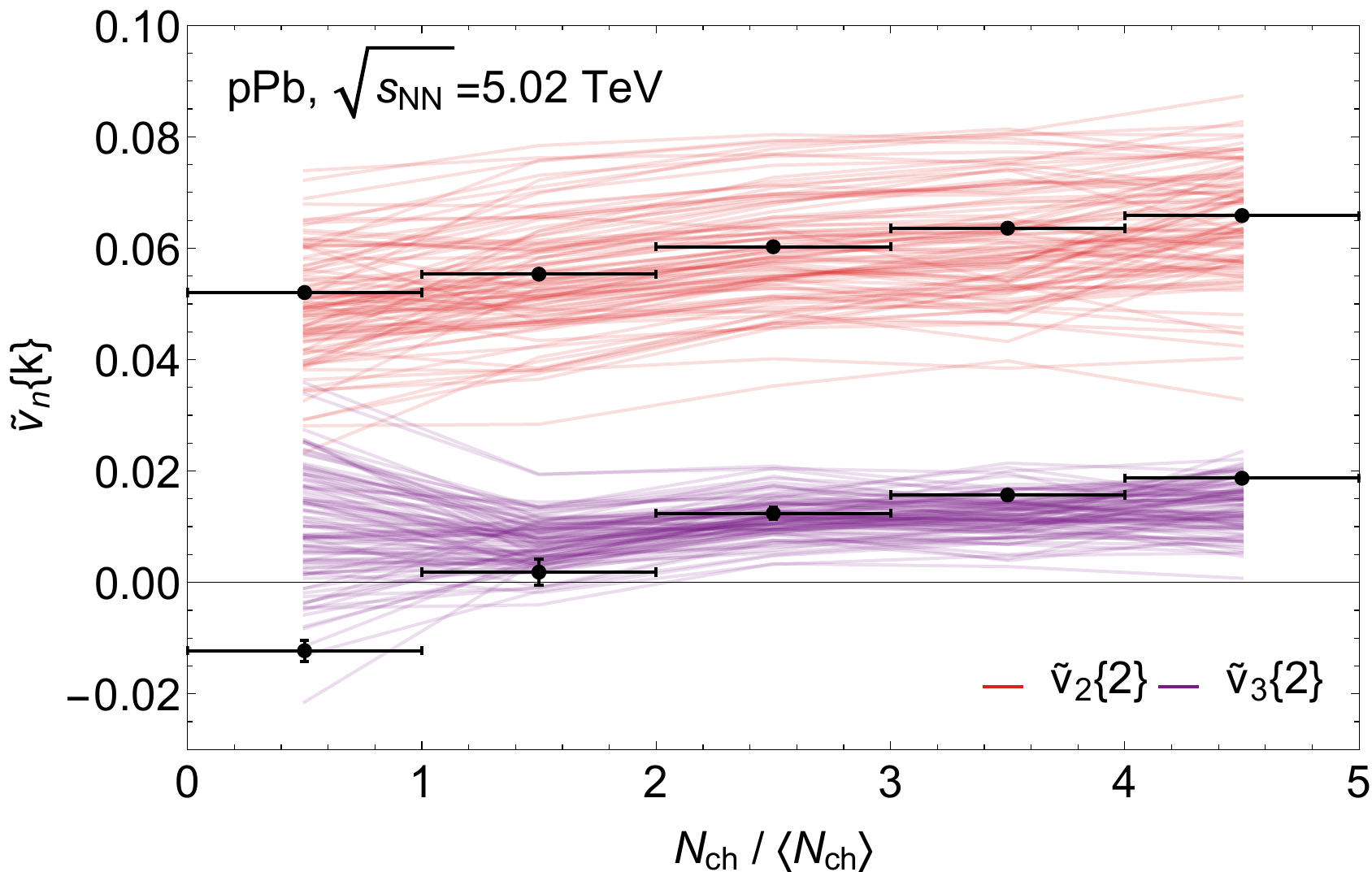}
\caption{\label{fig:postmultiplicity} A selection of the experimental data used together with 100 samples drawn from Fig.~\ref{fig:posterior1d}, with at (top) Multiplicities and transverse energy versus centrality (left), $p_T$ spectra for pions, kaons and protons for 0-5\% (solid) and 40-50\% (dashed,~*0.1)  centrality classes (middle) and $v_2\{2\}$ versus $p_T$ for 0-5\% (dashed) and 20-30\% (solid).\newline
(bottom) $v_n\{k\}$ versus centrality for PbPb collisions at top LHC energy (left), mean transverse momenta for $p$Pb collisions versus centrality (middle) and $\tilde v_n\{k\}$ %
for $p$Pb collisions depending on multiplicity class (right).%
}
\end{figure*}

For the hydrodynamic evolution, we solve the conservation equations for the stress-energy tensor, with the stress-energy tensor given by the hydrodynamic constitutive relation:
$T^{\mu\nu} = \rho u^\mu u^\nu - (P + \Pi)\Delta^{\mu\nu} + \pi^{\mu\nu},$
where $\Delta^{\mu\nu} = g^{\mu\nu} - u^\mu u^\nu$, and we use the mostly minus convention for the metric.
Here $\rho$, $P$, $\Pi$ and $\pi^{\mu\nu}$ are the energy density, pressure, bulk viscous pressure and the traceless transverse shear stress respectively.
The equations of motion for the bulk pressure $\Pi$ and the shear stress $\pi^{\mu\nu}$ are given by the 14-moment approximation \cite{Denicol:2014vaa}, where we keep only the transport coefficients used in \cite{Bernhard:2019bmu}:
\begin{align}
& D\Pi = -\frac{1}{\tau_\Pi}\left[\Pi + \zeta\nabla\cdot u + \delta_{\Pi\Pi}\nabla\cdot u\Pi - \lambda_{\Pi\pi}\pi^{\mu\nu}\sigma_{\mu\nu}\right],\nonumber\\
& \Delta^\mu_\alpha\Delta^\nu_\beta D\pi^{\alpha\beta} = -\frac{1}{\tau_\pi}\large[\pi^{\mu\nu} - 2\eta\sigma^{\mu\nu} + \delta_{\pi\pi}\pi^{\mu\nu}\nabla\cdot u\nonumber\\
& \quad - \phi_7\pi_\alpha^{\langle\mu}\pi^{\nu\rangle\alpha} + \tau_{\pi\pi}\pi_\alpha^{\langle\mu}\sigma^{\nu\rangle\alpha} - \lambda_{\pi\Pi}\Pi\sigma^{\mu\nu}\large].\label{eq:intro:secondordershear}
\end{align}
The pressure is given in terms of the energy density by the hybrid HotQCD/HRG equation of state \cite{Huovinen:2009yb,Bazavov:2014pvz,Bernhard:2018hnz}\@.
We parameterize the first order transport coefficients $\eta$ and $\zeta$ in terms of the dimensionless ratios $\eta/s$ and $\zeta/s$\@. 
In particular, $\eta/s = a + b (T-T_c)(T/T_c)^c$, with a minimal value $a = (\boldsymbol{\eta}/\bf{s})_\text{min}$ at $T_c = 154\,\text{MeV}$, a slope $b = (\boldsymbol{\eta}/\bf{s})_\text{slope}$ and a curvature $c = (\boldsymbol{\eta}/\bf{s})_\text{crv}$\@.
The bulk viscosity $\zeta/s$ is described by an unnormalized Cauchy distribution with height $(\boldsymbol{\zeta}/\bf{s})_\text{max}$, width $(\boldsymbol{\zeta}/\bf{s})_\text{width}$ and peak temperature $(\boldsymbol{\zeta}/\mathbf{s})_{T_0}$\@.
The second order transport coefficients $\tau_\Pi$, $\delta_{\Pi\Pi}$, $\lambda_{\Pi\pi}$, $\tau_\pi$, $\delta_{\pi\pi}$, $\phi_7$, $\tau_{\pi\pi}$ and $\lambda_{\pi\Pi}$ are also given in terms of dimensionless ratios.
Of these, we fix
\[
\frac{\delta_{\Pi\Pi}}{\tau_\Pi} = \frac{2}{3}, \ \ \frac{\lambda_{\Pi\pi}}{\tau_\Pi\delta} = \frac{8}{5}, \ \ \frac{\delta_{\pi\pi}}{\tau_\pi} = \frac{4}{3}, \ \ \phi_7P = \frac{9}{70}, \ \ \frac{\lambda_{\pi\Pi}}{\tau_\pi} = \frac{6}{5}
\]
to the values from kinetic theory \cite{Denicol:2014vaa}, with $\delta=1/3-c_s^2$, while we vary the shear and bulk relaxation times $\tau_\pi$ and $\tau_\Pi$ as well as one other second order coefficient $\tau_{\pi\pi}$. We vary these according to the ratios
\[
\frac{\boldsymbol{\tau}_\Pi\mathbf{sT}\boldsymbol{\delta}^2}{\boldsymbol{\zeta}}, \ \ \frac{\boldsymbol{\tau}_\pi\mathbf{sT}}{\boldsymbol{\eta}}\,{\rm  and } \ \ \frac{\boldsymbol{\tau}_{\pi\pi}}{\boldsymbol{\tau}_\pi}.
\]

Finally, the hydrodynamic fluid undergoes particlization at a temperature $\bf{T_{\rm switch}}$, whereby viscous contributions as well as resonances are included according to the algorithms presented in \cite{pratt:2010jt,Bernhard:2018hnz}. These hadrons are then evolved using the SMASH hadronic cascade code \cite{Weil:2016zrk,dmytro_oliinychenko_2020_3742965,Sjostrand:2007gs}\@.

\noindent
{\bf Experimental data -}
To compare our model to experiment we start with the dataset used in \cite{Bernhard:2016tnd}: PbPb charged particle multiplicity $dN_{ch}/d\eta$ at 2.76 \cite{Aamodt:2010cz} and 5.02 TeV \cite{Adam:2015ptt}, transverse energy $dE_T/d\eta$ at 2.76 TeV \cite{Adam:2016thv}, identified yields $dN/dy$ and mean $p_T$  for pions, kaons and protons at 2.76 TeV \cite{Abelev:2013vea}, integrated anisotropic flow $v_n\{k\}$ for both 2.76 and 5.02 TeV \cite{Adam:2016izf} and $p_T$ fluctuations \cite{Abelev:2014ckr} at 2.76 TeV\@. On top of this we added identified transverse momentum spectra using six coarse grained $p_T$-bins separated at $(0.5, 0.75, 1.0, 1.4, 1.8, 2.2, 3.0)\,$GeV both for PbPb at 2.76 \cite{Abelev:2013vea} and $p$Pb at 5.02 TeV \cite{Adam:2016dau}, anisotropic identified flow coefficients using the same $p_T$ bins (statistics allowing) at 2.76 \cite{Adam:2016nfo} and 5.02 TeV \cite{Acharya:2018zuq}. As in \cite{Moreland:2019szz} we use $\tilde v_n\{k\}$ anisotropic flow coefficients for $p$Pb at 5.02 TeV \cite{Aaboud:2017acw} \footnote{Since in $p$Pb $v_n\{k\}$ can become imaginary we use $\tilde v_n\{k\} \equiv {\rm sgn}(v_n\{k\}^k) |v_n\{k\}|$.} as well as mean $p_T$ for pions, kaons and protons at 5.02 TeV  \cite{Abelev:2013haa}. All of these use representative centrality classes, whereby we also specifically included high multiplicity $p$Pb classes for its anisotropic flow coefficients, giving a total of 418 and 96 datapoints for PbPb and $p$Pb collisions respectively.

\noindent
{\bf Posterior distribution -}
In order to estimate the likelihood of all 21 parameters (bold in the model) we used \emph{Trajectum} \footnote{Source code is available at \url{https://sites.google.com/view/govertnijs/trajectum}.} to simulate the full PbPb ($p$Pb) model at 1000 (2000) design points located on a Latin Hypercube in the parameter space using 6k (40k) events per design point (the parameter ranges can be found in the posterior distributions later) \footnote{The computing time for \emph{Trajectum} and SMASH are comparable, both taking roughly $10^6$ core hours.}. For each system we apply a transformation to 25 principal components (PCs), for which we train Gaussian emulators \cite{williams2006gaussian,Bernhard:2018hnz,Moreland:2019szz}. Crucially, the emulator also estimates its own uncertainty (which we validated) and through the Principle Component Analysis this includes correlations among the datapoints.
Full details as well as emulator results can be found in our companion paper \cite{Bayesianlong}.

\begin{figure}[ht]
\includegraphics[width=0.49\textwidth]{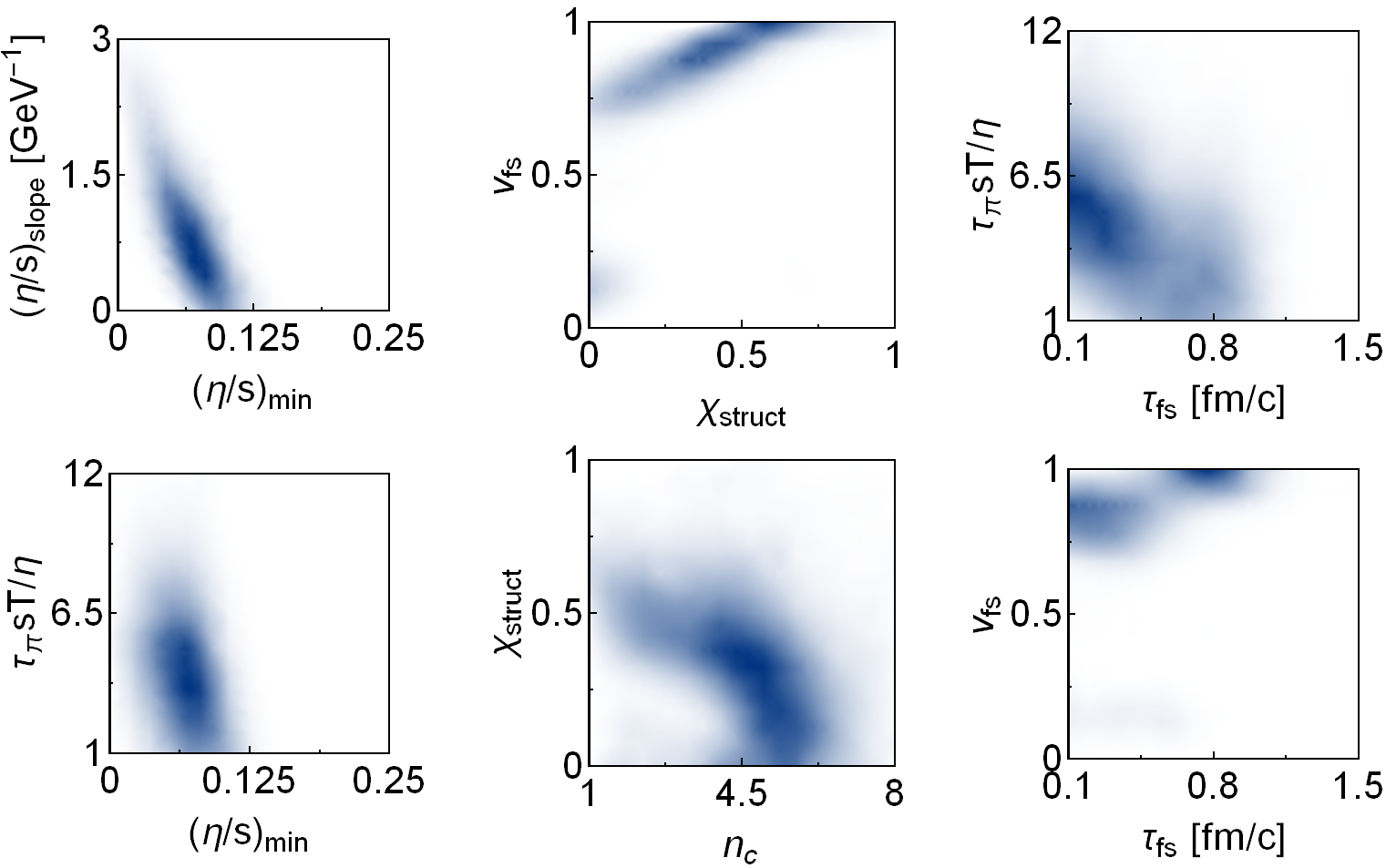}
\caption{\label{fig:posterior2d}We highlight a few interesting or strong correlations among the posterior distributions shown in Fig.~\ref{fig:posterior1d}.}
\end{figure}
\begin{figure}[ht!]
\includegraphics[width=0.235\textwidth]{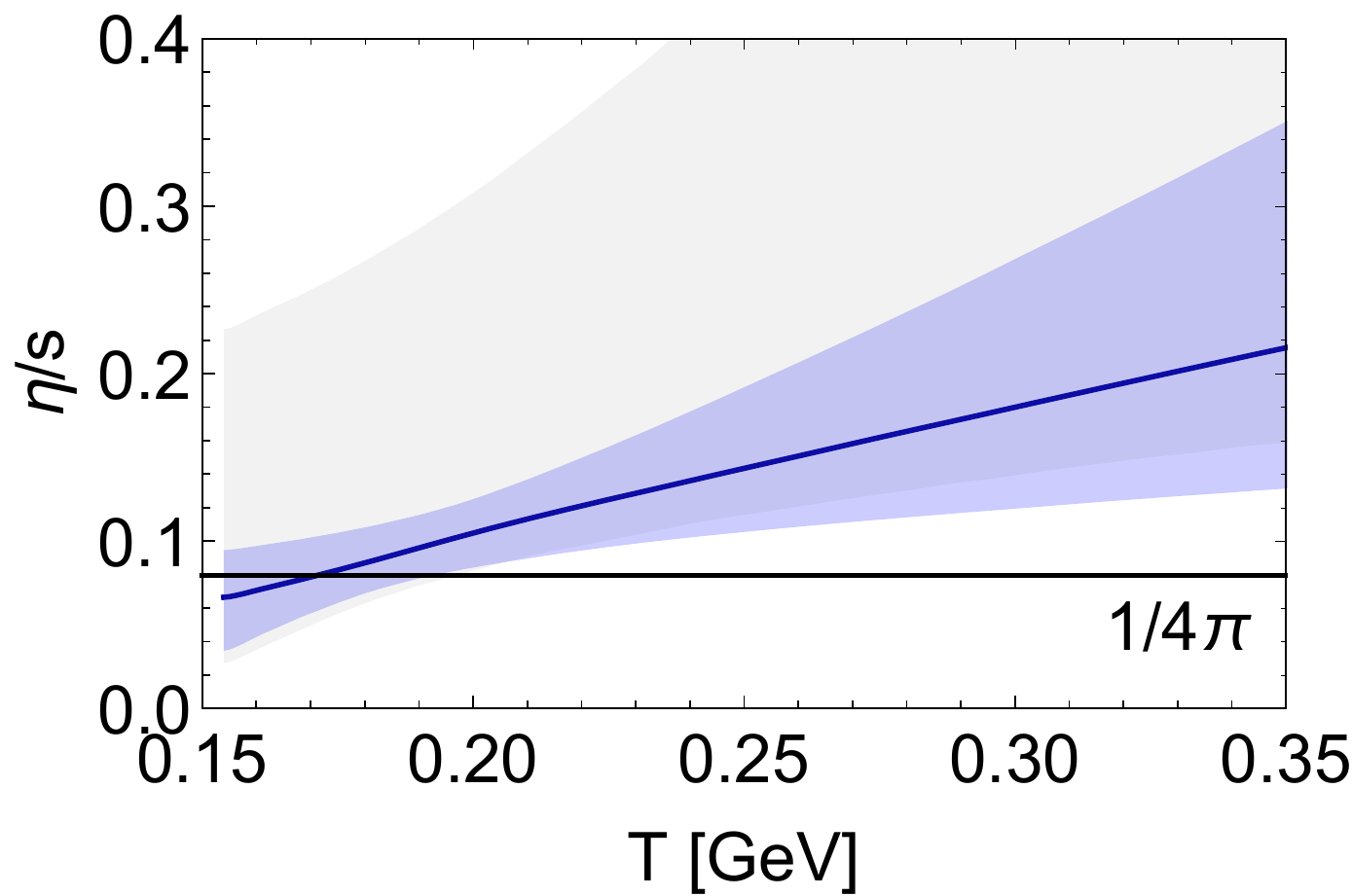}
\includegraphics[width=0.235\textwidth]{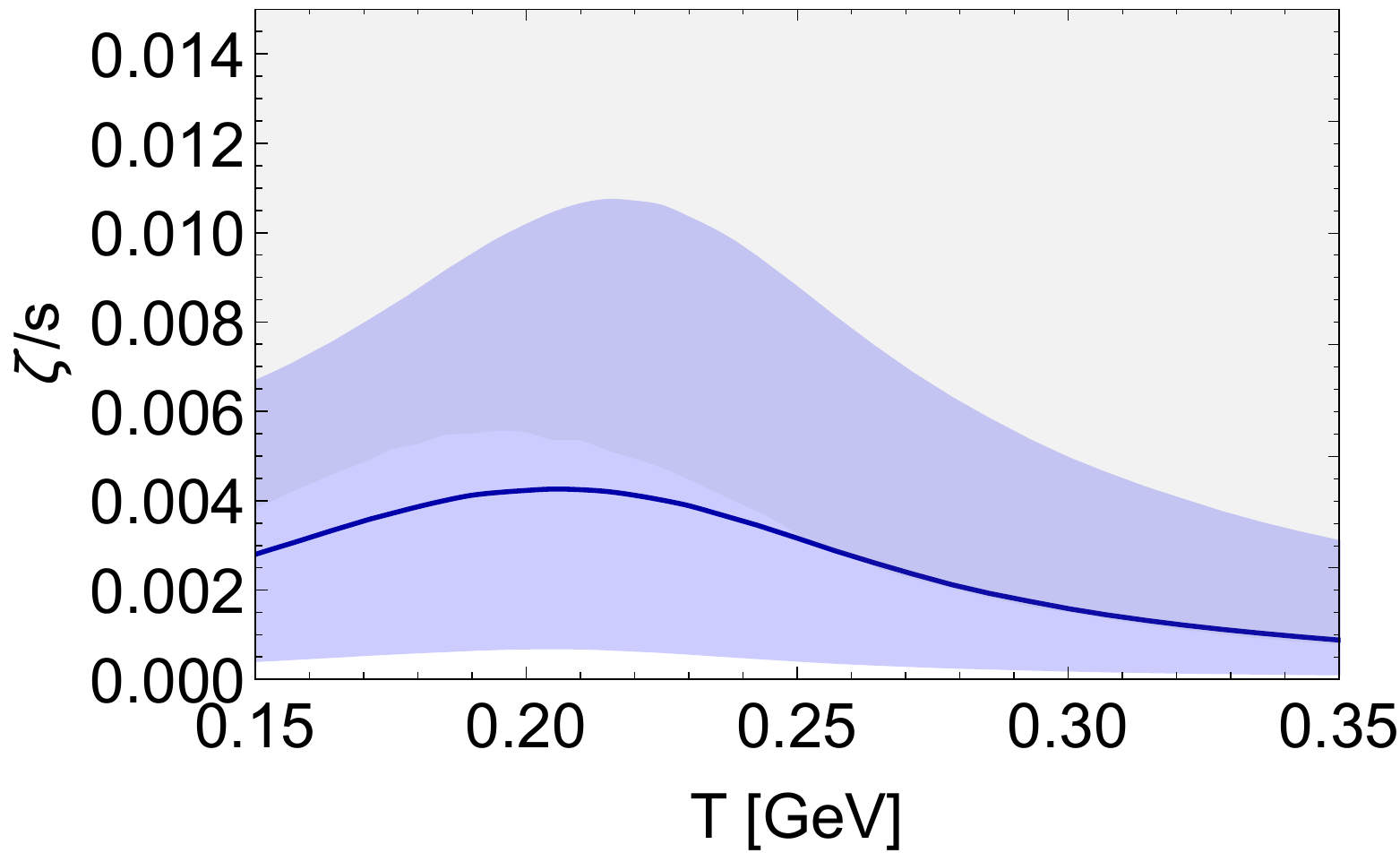}
\caption{\label{fig:viscosities}Posterior distributions for the specific shear and bulk viscosities versus temperature together with their mean and 90\% confidence band (blue). %
The 90\% confidence bands for the prior distribution is shown in gray (extending till 0.08 for $\zeta/s$, not shown).}
\end{figure}

Using either PbPb only or both PbPb and $p$Pb emulators we ran a Markov Chain Monte Carlo (mcmc) employing the EMCEE2.2 code \cite{ForemanMackey:2012ig,Bernhard:2018hnz,Moreland:2019szz}, using 600 walkers for approximately $15$k steps. This led to the converged posterior distributions in Fig.~\ref{fig:posterior1d}, shown with (solid) and without (dashed) the $p$Pb data. Fig.~\ref{fig:postmultiplicity} shows results from 100 random samples of the posterior distribution for a representative selection of our datapoints. In general these compare well, even for $p_T$-differentiated identified $v_n\{2\}$ distributions for both central and peripheral collisions.

For $p$Pb the posterior distributions are significantly wider than the experimental uncertainties, since even for 2000 design points the model is sufficiently complicated that a significant emulator uncertainty remains. It is for this reason that including $p$Pb for the posterior (blue solid versus green dashed in Fig.~\ref{fig:posterior1d}) does not change the probabilities as much as perhaps expected, though for parameters especially sensitive to small and short-lived systems better constraints are obtained ($\mathbf{n}_c$, $\boldsymbol{\tau}_{\rm fs}$, $\textbf{w}$, $\textbf{d}_{\rm min}$ and $\boldsymbol{\sigma}_{\rm fluct}$)\@.

\begin{figure*}[ht]
\includegraphics[width=0.99\textwidth]{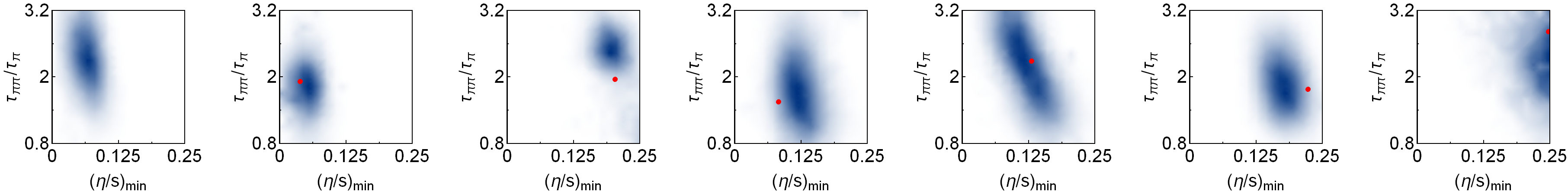}
\caption{\label{fig:correlation}We show an interesting physical correlation between $\boldsymbol{\tau}_{\pi\pi}/\boldsymbol{\tau}_\pi$ and $(\boldsymbol{\eta}/\bf{s})_\text{min}$ with posterior distributions fitted to experimental data (left) or six sets of generated data from random parameter settings (shown in red). All of these show a negative Pearson's correlation averaging -0.24.}
\end{figure*}

Perhaps the most striking feature in Fig.~\ref{fig:posterior1d} is that the posterior for the maximum of $\zeta/s$ peaks at zero. This is in contrast to previous work \cite{Ryu:2015vwa,Bernhard:2016tnd,Bernhard:2019bmu} that prefers a positive bulk viscosity in order to reduce the mean $p_T$. A larger bulk viscosity, however, makes it hard to describe the $p_T$ identified spectra (Fig.~\ref{fig:postmultiplicity} (middle,top)).

\begin{figure}[ht]
\includegraphics[width=0.9\columnwidth]{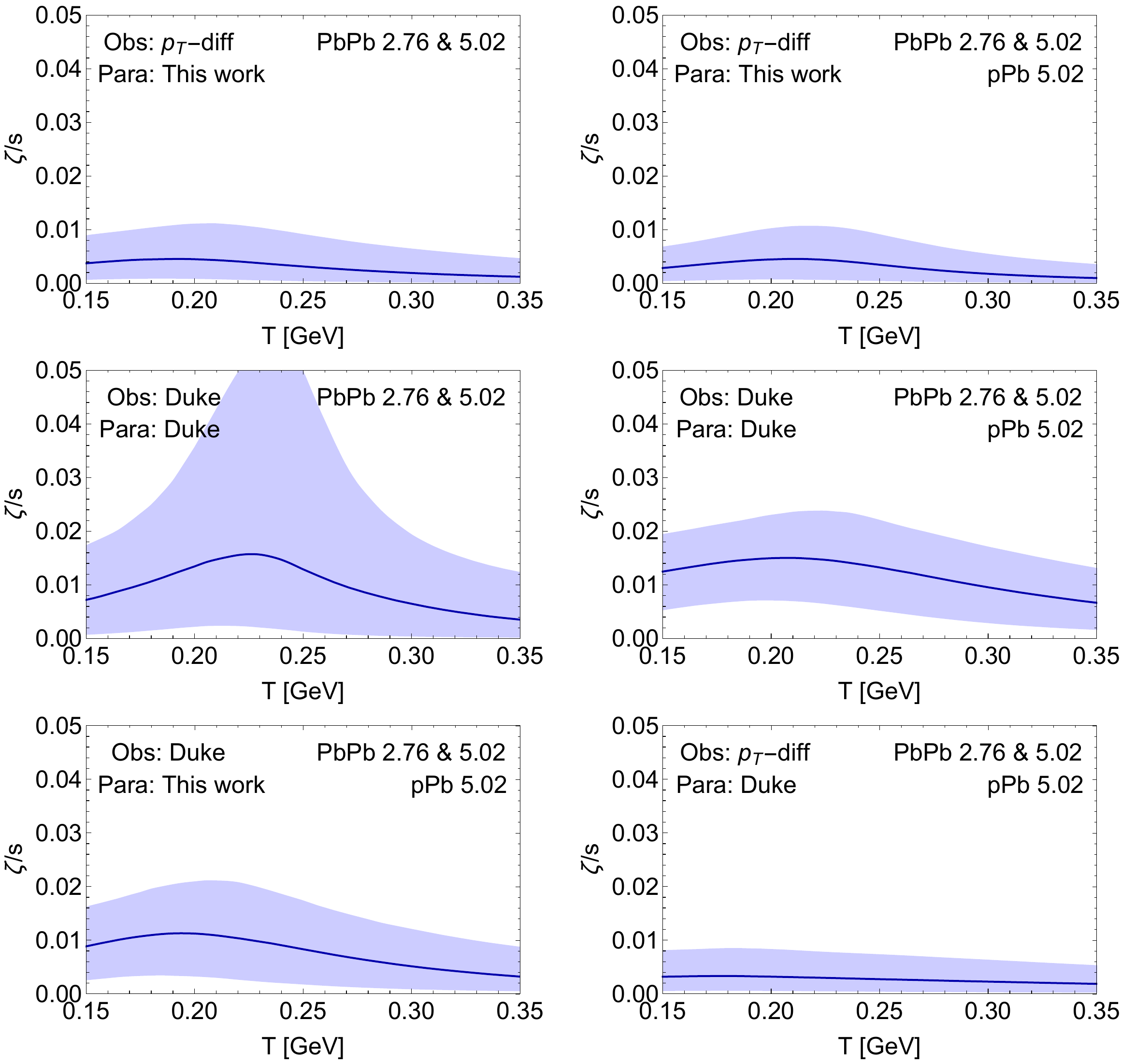}
\caption{\label{fig:bulkviscositiesmcmc}We show posteriors for $(\boldsymbol{\zeta}/\bf{s})_\text{max}$ as in Fig.~\ref{fig:viscosities} in variations using the model presented, or limiting to a subset with fewer observables (as in \cite{Bernhard:2016tnd}, labelled Duke), fewer parameters (as in \cite{Bernhard:2016tnd}, labelled Duke), or without the $p$Pb system. The new $p_T$-differential observables are the most significant addition that led to the small bulk viscosity in Fig.~\ref{fig:viscosities}.}
\end{figure}

Given the scope of our 21 parameter model it is perhaps not surprising that constraints on the parameters and in particular the second order transport coefficients are not that strong. We do however see interesting correlations, as shown in Fig.~\ref{fig:posterior2d}. 
As expected $(\boldsymbol{\eta}/\bf{s})_\text{min}$ and $(\boldsymbol{\eta}/\bf{s})_\text{slope}$ are negatively correlated.
Perhaps the most interesting correlation is between $\boldsymbol{\tau}_{\rm fs}$ and $\boldsymbol{\tau}_\pi \bf{sT}/\boldsymbol{\eta}$: it is possible to have a rather long free streaming time, but only if $\tau_\pi$ is relatively small. This correlation indeed guarantees the quick applicability of  (viscous) hydrodynamics.
In the pre-equilibrium stage, the larger $\boldsymbol{\tau}_\text{fs}$ the larger the deviation from hydrodynamics, see also \cite{Liu:2015nwa, Bayesianlong}\@.
On the other hand, a shorter  shear relaxation time  dampens large deviations from viscous hydrodynamics more quickly.
In this way, we can interpret the joint constraint on $\boldsymbol{\tau}_\pi \bf{sT}/\boldsymbol{\eta}$ and $\boldsymbol{\tau}_\text{fs}$ from the posterior distribution as a preference for the fluid to quickly hydrodynamize \cite{Heller:2016gbp}\@.
Another strong negative correlation is between $\bf{v}_\text{fs}$ and $\boldsymbol{\chi}_\text{struct}$, and indeed for ${\bf v}_\text{fs} = 1$ our $\boldsymbol{\chi}_\text{struct}$ distribution is in agreement with \cite{Moreland:2018gsh}.
This highlights the importance of gaining a better understanding of the initial stages of the collision.
We also find that $\bf{n}_c$ and $\boldsymbol{\chi}_\text{struct}$ are negatively correlated.

\begin{figure}[ht]
\includegraphics[width=0.99\columnwidth]{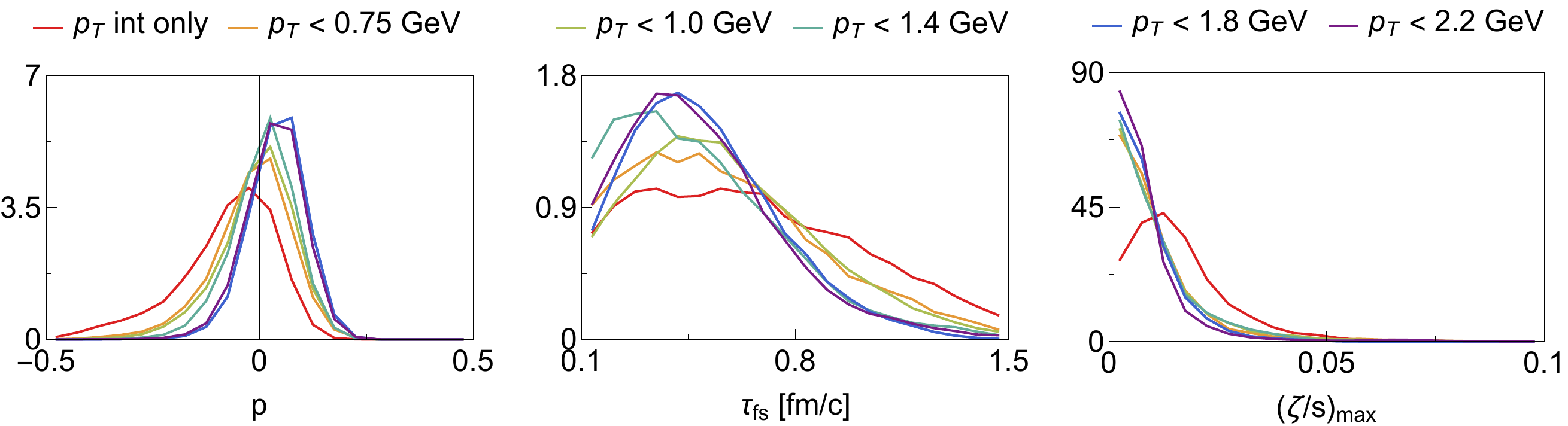}
\caption{\label{fig:ptcut}We vary the maximum $p_T$ used in the $p_T-$differentiated analysis for three representative parameters. Clearly for the bulk viscosity the highest gain in precision is present by including low $p_T$ particles, whereas the Trento parameter and free streaming time benefit more gradually from higher $p_T$ data.}
\end{figure}

We obtain tight constraints on our prehydrodynamic flow parameter $\bf{v}_{\rm fs}$. 
Indeed the preferred value is close to unity, with perhaps an unlikely option of a small velocity combined with a tiny $\boldsymbol{\chi}_\text{struct}$.
From Fig.~\ref{fig:posterior2d} we see that either there is a short $\boldsymbol{\tau}_\text{fs}$ with ${\bf v}_{\rm fs}\approx 0.85$, or a longer $\boldsymbol{\tau}_\text{fs}$ with ${\bf v}_{\rm fs}=1$. The first option is consistent with an equation of state that is slightly below the conformal limit; indeed at $T=0.4\,$GeV the pressure equals 85\% of its conformal value \cite{Huovinen:2009yb,Bernhard:2018hnz}. In this scenario the fluid is initialized with a bulk pressure close to its ideal hydrodynamic value at an early time (see also \cite{Bayesianlong}). In the second scenario the fluid will start further from equilibrium, but here the influence of $\bf{v}_{\rm fs}$ on the initial geometry is dominant, as is also clear from the correlation with $\boldsymbol{\chi}_\text{struct}$\@.

A closure test is crucial for any model that attempts parameter extractions as ours. For this we extracted posterior distributions for generated `data' sets at six random points in our parameter space. Apart from confirming the probability distributions this can also lead to physical insights of wider applicability than by just using the true experimental data. An example is shown in Fig~\ref{fig:correlation} showing that often $(\boldsymbol{\eta}/\bf{s})_\text{min}$ correlates negatively with $\boldsymbol{\tau}_{\pi\pi}/\boldsymbol{\tau}_{\pi}$\@. The interpretation is that both have similar effects on the elliptic flow. Both the shear viscosity and $\tau_{\pi\pi}$ are dissipative, and hence tend to isotropize the plasma, which suggests an explanation for this correlation (see also the correlation between $(\boldsymbol{\eta}/\bf{s})_\text{min}$ and $\boldsymbol{\tau}_\pi \bf{sT}/\boldsymbol{\eta}$ in \Fig{fig:posterior2d}).

\noindent{\bf Discussion -}
From the posterior distributions it is possible to obtain the 90\% confidence limit of the viscosities, as shown in Fig.~\ref{fig:viscosities}. 
For lower temperatures the viscosity is consistent with the canonical string theory value of $1/4\pi$ \cite{Policastro:2001yc} (note also that stringy models exist with a lower viscosity \cite{Brigante:2007nu}). 
There is a clear tendency for $\eta/s$ to increase, as expected from the running of the coupling constant. As already shown the bulk viscosity is found to be small, which is consistent with an approximately conformal theory.
This contrasts with the prevailing view that a finite bulk viscosity \cite{Ryu:2015vwa,Bernhard:2016tnd,Bernhard:2019bmu} is needed in order to simultaneously fit the mean transverse momenta and anisotropic flow.
We also obtain mild constraints on $\boldsymbol{\tau}_\pi {\bf sT}/\boldsymbol{\eta} \lesssim 7$ and $\boldsymbol{\tau}_{\pi\pi}/\boldsymbol{\tau}_\pi \gtrsim 1.5$. The value for $\boldsymbol{\tau}_\pi \bf{sT}/\boldsymbol{\eta}$ compares well with the holographic  ($4-\log(4)\approx 2.61$, \cite{Baier:2007ix}) and weak coupling (5, \cite{Denicol:2014vaa}) results. The  $\boldsymbol{\tau}_{\pi\pi}/\boldsymbol{\tau}_\pi$ value is consistent with the weak coupling result ($10/7 \approx 1.43$, \cite{Molnar:2013lta}) and agrees well with the holographic result ($88/35(2-\log(2))\approx 1.92$, \cite{Bhattacharyya:2008mz})\@. 

We performed several extra mcmc analyses in order to better understand our small bulk viscosity, shown in Fig.~\ref{fig:bulkviscositiesmcmc}. 
There we varied the observables used (all versus those  in \cite{Bernhard:2016tnd} labelled Duke) and similarly our varying parameters and lastly including $p$Pb or not. Setting both observables, parameters and systems to those in \cite{Bernhard:2016tnd} reproduces their bulk viscosity. Including $p$Pb, more parameters and, most importantly, including our $p_T$-differential observables all reduce the bulk viscosity, explaining the result in Fig.~\ref{fig:viscosities}. The fact that setting our parameters to the ones used in \cite{Bernhard:2016tnd} reproduces the small bulk viscosity in particular implies that setting $v_{\rm fs}=1$ does not lead to a larger bulk viscosity, but instead gives a smoother sub-nucleonic structure as is clear from the $v_{\rm fs}$-$\chi_{\rm struct}$ correlation in Fig.~\ref{fig:posterior2d}.

A crucial question on our analysis is how much information is gained by the respective $p_T$ bins and how sensitive our results are to the observables at high $p_T$. This is particularly important, since our viscous freeze-out prescription  \cite{pratt:2010jt,Bernhard:2018hnz} has significant systematic uncertainty that is more important at high $p_T$ (see also \cite{Bernhard:2016tnd}). It is hence important to verify that our main conclusions are not sensitive to our particular freeze-out prescription, and indeed we see in Fig.~\ref{fig:ptcut} that the bulk viscosity is almost entirely determined by the low $p_T$ bins. For other observables such as the free streaming time a more gradual increase in precision is observed, but none of our posteriors depend sensitively on our highest $p_T$ bin between $2.2 - 3.0\,$GeV.

It is still debated whether matter formed in $p$Pb collisions can be described by hydrodynamics \cite{Weller:2017tsr,Kurkela:2019kip,Nagle:2018nvi}. Indeed, one of our main motivations of this study was to shed light on this question, by computing posterior probabilities with and without $p$Pb collisions. If our framework manages to fit $p$Pb well this gives further evidence for a hydrodynamic picture. 
In general we agree well with $p$Pb observables, but the mean kaon $\langle p_T\rangle$ seems to deviate significantly (see also \cite{Bayesianlong}). This can either imply a deviation from the hydrodynamic picture, but given that our $p$Pb observables are much harder to emulate it could also indicate a more advanced analysis within hydrodynamics or a more advanced initial stage is needed.

Our model can be improved in two directions. Firstly, our initial state, prehydrodynamic phase and particlization are not based on a microscopic theory and in particular the transition to hydrodynamics is not smooth \cite{vanderSchee:2013pia}. It would be interesting to investigate this point by including a more physically motivated initial stage. Secondly, we were only able to use data that can be reliably estimated using about 6k events, whereas much more sophisticated data is available. Our data includes the widest set available for a global analysis so far, but nevertheless only roughly 20 principal components are non-trivial. This is much more than in e.g.~\cite{Bernhard:2018hnz,Moreland:2018gsh,Auvinen:2020mpc} where up to 8 PCs contain over 99\% of the non-trivial information. Nevertheless the question remains whether it is possible to estimate so many parameters with relatively limited experimental data (see also \cite{Everett:2020yty} where it is found that using only the limited dataset it is difficult to obtain much stronger constraints than the given prior range).
In the future it will hence be important to incorporate more non-trivial data, perhaps using some approximations  to reduce computation time (see also \cite{Bayesianlong,Niemi:2015qia}). %

{\bf Acknowledgements -} We are grateful to Jonah Bernhard and Scott Moreland for making their codes public together with an excellent documentation. We thank Steffen Bass, Aleksas Mazeliauskas, Ben Meiring and Urs Wiedemann for discussions.
GN is supported by the U.S. Department of Energy, Office of Science, Office of Nuclear Physics under grant Contract Number DE-SC0011090.

\bibliography{letter, manual}
\end{document}